\documentstyle[12pt]{article}
\newcommand{\zero}{\setcounter{equation}{0}}

\begin{document}

\begin{center}

{\Large \bf Screwed superconducting cosmic strings}

C.N. Ferreira$^*$ \footnote{crisnfer@cbpf.br}

$^*${\it Centro Brasileiro de Pesquisas F\'{\i}sicas, Rua Dr.
Xavier Sigaud 150, Urca 22.290-180, Rio de Janeiro, RJ, Brazil}

\end{center}



\begin{abstract}
We show that it is possible to build up a consistent model
describing a superconducting cosmic string (SCS) endowed with
torsion. A full string solution is obtained by matching the
internal and the external solutions. We derive the deficit angle,
the geodesics of and the gravitational force on a test-particle
moving under the action of this screwed SCS. A couple of potentially observable
astrophysical phenomena are highlighted: the dynamics of compact
objects orbiting torsioned SCS and the accretion of matter onto it.
\end{abstract}

\newpage


\section{Introduction}

Cosmic strings are a class of solutions \cite{Vilenkin} which
represent topological defects that may have been formed during
phase transitions in the realm of the Early Universe
\cite{Kibble}. The GUT defects carry a large energy density and,
hence, are of interest in Cosmology, as potential sources for
primordial density perturbations\cite{avelino98}. These
fluctuations would leave their imprint in the cosmic microwave
background radiation (CMBR); a prediction not fully ruled out by
COBE satellite observations yet\cite{smoot99,Pogosian}; so, 
they would act as seeds for structure formation and consequently  builders of
the largest-scale structures in the Universe\cite{Brandenberger},
such as the very high redshift superclusters of galaxies (as for
instance the {\it great wall}). They may also help to explain the
most energetic events in the Universe, like  the cosmological
gamma-ray bursts (GRBs)\cite{brandeb93}, ultra high energy cosmic
rays (UHECRs) and very high energy neutrinos\cite{Brandenberger},
and gravitational-wave bursts\cite{HD00} and
backgrounds\cite{Allen94}. All these are issues deserving
continuous investigation by many physicists nowadays
\cite{Hawking}.

Witten \cite{Witten1} has shown that cosmic strings may possess
superconducting properties and may behave like both  bosonic (see
Ref.[\cite{Patrick1}] and references therein) and fermionic
strings \cite{Jackiw,Davis99}. The relevant superconductivity is
generated during, or very soon after, the primary phase transition
in which the string was formed.

A further question of much relevance concerns the consideration of gravitational
effects on the formation of SCS's. In this direction, it has already been 
discussed by several authors that   torsion
may have been an important element in the early Universe, when the
quantum effects of gravitation were drastically
important\cite{Sabbata1,Yishi}.

Our present understanding of the early Universe and the Planck era
indicates that extended objects, like strings, are the best framework to describe the
physics between the GUT and the Planckian scales \cite{Deser1}. String
corrections to Quantum Gravity\cite{Boulware} lead to effective
gravitational models where torsion plays a very relevant r\^ole
through $\alpha'$-corrections terms involving curvature
and torsion \cite{Deser2,Shapiro2}. So, our viewpoint in
this paper is that torsion effects are to be taken into account and
are supposed to persist till the cosmic strings were formed.

The Universe in its very early era
($10^{18} GeV \geq T \geq 10^{16} GeV $, the era before the GUT
phase transition ) consisted of torsion
particles and GUT (SU(5)) particles (heavy fermions, gauge bosons
and Higgs particles); torsion particles couple with intrinsic spins of matter.
Therefore, this interaction appears in the era in which the
gravitational interaction became important for microscopic
systems, such as the very early universe\cite{Ogino}.

In \cite{Yishi}, to describe the space-time defects \cite{Ross,Sabbata3} in the early
universe in an invariant form, the authors adopt  a new topological
invariant, obtained by means of the torsion tensor, 
to measure the size of defects and interpret them as
the quantized dislocation flux in internal space.  Related to these arguments,
it may become a relevant issue to contemplate topological defects described by
Superconducting Cosmic Strings (SCS) coupled torsion;
they  may be present as a relic in ours days.

Cartan torsion has been  previously related to  spinning cosmic
strings \cite{Soleng92,Letelier1} from quite distinct points of
view. In particular,  Soleng\cite{Soleng92} has studied the case
in which the cosmic string can be identified with a cylinder
filled in with an anisotropic fluid possessing  an intrinsic spin
polarized along the string symmetry axis. It was shown there that
the spin angular momentum generates torsion as a natural
consequence of extending Einstein's theory to include
 Riemann-Cartan geometry. Such a spinning string exhibits interesting gravitational
 effects, including inertial dragging and the possible generation  of closed time-like
 lines.\cite{Soleng92} Although that theory was consistently developed, one has not
 discussed neither the astrophysical nor cosmological constraints on the potential implications
 of its realization in Nature. Moreover, as it is easy to see from the
Arkuszewski-Kopczy\'nski-Ponomariev matching conditions (Eq.18 in
Ref.\cite{Soleng92}), the spin (i.e., torsion) effects do not
propagate outwards. This is, of course, due to the fact that the
string is composed of an anisotropic spin-polarized fermionic
fluid. Nonetheless, this property (spin effects propagation)
manifests itself as a space-time metric, as stressed by the
definition (coupling) introduced in the Eq.(6) of Soleng's
paper\cite{Soleng92}. Conversely, our model does exhibit
propagation of torsion effects outwards, and this may be
responsible for the appearance of new, unexpected phenomena, as
discussed below. Far-reaching effects are possible because, to
obtain a complete string solution, we need to introduce
superconductivity, which allows us to relate the torsion field of our
model to the vortex characterizing the defect.

Recently, Kleinert has pointed out a number of aspects and effects
of torsion\cite{Kleinert,Kleinert1,Kleinert2}. In his work,
the gravitational counterpart  yields consistent results
only for completely antisymmetric or gradient torsion.
If one takes into account that even spinless
particles experience a torsion-originated force, one might
naturally expect them to be also source for torsion. In a line of
works related to physical consequences of torsion, Kleinert
shows that deviations from Einstein gravity's effects may be
attained if one adds to the gravity action a gradient term for
torsion. According to these works, a good consistency test for
gravity with torsion is the criterium that the torsion coupling to
matter must be such that spinless particles run on autoparallel,
rather than geodesic, trajectories.
On the other hand, it is to be noticed that massless and
massive gauge vector gauge bosons couple differently to torsion.
To accommodate this fact with the mass generation via a Higgs
mechanism, one avoids inconsistencies if the Higgs scalar runs
along autoparallel trajectories.

In this work, we study the  case of bosonic SCS's in Riemann-Cartan
space-time with coupling terms in the potential driving the string
dynamics.  We aim at dealing with more realistic models which may
demand supersymmetry, an essential ingredient for Grand-Unified
theories and string theories. Thus, we ought to combine both
gravitational and spin degrees of freedom in the same formalism;
thus, torsion is required.

The main-stream of this paper is as follows: we explore the
physics of torsion coupling to cosmic strings in Section II. An
external solution for the SCS metric in this scenario
in presented in Section III, while in Section IV we derive
the corresponding metric for the internal structure of
the SCS by using the weak-field approximation. The complete
solution of the SCCS is obtained in Section V, by using the
joint conditions. In section VI, we obtain
a neat expression for the deficit angle in this context. Two
applications are provided in Section VII. It is shown that such a
high intensity of the gravitational force from the screwed SCS
(when compared with the one generated by a superconducting string)
may have important effects on the dynamics of compact objects
orbiting around it, and also on matter being accreted by the
string itself.

\section{Torsion coupling to cosmic strings \zero}

Here, we propose a consistent framework for the torsion field
pervading a cosmic string
and define the vortex configuration for this problem. We choose here
to analyze the simplest case where  torsion appears. In this line
of reasoning, it is possible to describe torsion as a gradient-like
field \cite{Kleinert}

\begin{equation} S_{\mu \nu}^{\hspace{.3 true cm} \lambda}
=\frac{1}{2} [\delta_{\mu}^{\lambda} S_{\nu} -
\delta_{\nu}^{\lambda} S_{\mu} ], \label{torc1}
\end{equation}

\noindent
the affine connection being written as

\begin{equation}
 \Gamma_{\lambda \nu}^{\hspace{.3 true cm} \alpha} =
\{^{\alpha}_{\lambda \nu}\} +
 S^{\alpha}g_{\lambda \nu} - S_{\lambda} \delta^{\alpha}_{\nu}, \label{kont1}
\end{equation}

\noindent
where $S_\alpha = \partial_{\alpha} \Lambda$ is the only piece contributing
to torsion, given in Eq.(\ref{torc1}).  Here, the $\Lambda$-field is
the source of torsion in the string.

We may consider a theory of {\it gravitation} possessing
torsion by writing the part of the action $I$ stemming from the
curvature scalar $R$ as:

\begin{equation}
I = \int d^4x \sqrt{g} \left[\frac{1}{16 \pi
G}R(\{\}) - \frac{\alpha}{2} \partial_{\mu} \Lambda
\partial^{\mu} \Lambda \right] + I_m, \label{acao1}
\end{equation}

\noindent
where $R(\{\})$ is the curvature scalar of the Riemannian theory and
$I_m$ is the matter action which describes the superconducting cosmic
string. The coupling constant $\alpha $ will be specified with the help
of COBE data.

We can study the SCS considering the Abelian Higgs model with two scalar
fields, $\phi$ and $\tilde \Sigma $. In this case, the action for all matter
fields turns out to be:

\begin{equation}
I_m = \int d^4x \sqrt{{g}}[-\frac{1}{2}D_{\mu}\phi
(D^{\mu}\phi)^* - \frac{1}{2}D_{\mu} \Sigma (D^{\mu}
\Sigma)^* - \frac{1}{4}F_{\mu \nu}F^{\mu \nu} - \frac{1}{4}H_{\mu
\nu}H^{\mu \nu} - V(|\phi|, |\Sigma |, \Lambda)],
\end{equation}

\noindent
where $ D_{\mu} \Sigma = (\partial_{\mu} + ie A_{\mu}) \Sigma $ and
$D_{\mu} \phi = ( \partial_{\mu} + i qC_{\mu})\phi $ are the covariant
derivatives. The reason why the gauge fields do not minimally couple to
torsion is well discussed in Refs.\cite{Gaspperini,Hehl}.  The field
strengths are defined as usually as $ F_{\mu \nu} =
\partial_{\mu}A_{\nu} - \partial_{\nu} A_{\mu}$ and $ H_{\mu \nu} =
\partial_{\mu}C_{\nu} - \partial_{\nu} C_{\mu} $, with $A_\mu$ and
$C_\nu$ being the gauge fields.

The potential $V(\varphi, \sigma, \Lambda)$ triggering the
symmetry breaking can be fixed by:

\begin{equation} V(\varphi, \sigma, \Lambda) = \frac{\lambda_{\varphi}}{4} (
\varphi ^2 - \eta^2)^2 + f_{\varphi \sigma}\varphi ^2\sigma ^2 +
\frac{\lambda_{\sigma}}{4}\sigma ^4 -
\frac{m_{\sigma}^2}{2}\sigma^2 + l^2\sigma^2 \Lambda^2, \end{equation}

\noindent
where $\lambda_{\varphi}$, $\lambda_{\sigma}$, $f_{\varphi \sigma}$, $m_{\sigma}$ and $l^2$
are coupling constants. Constructed this manner, this potential possesses all the ingredients that make it viable the formation of a superconducting cosmic string, as it is well-stablished. In addition, it is extended to include a new term describing the interaction with the torsion field. The presence of this interaction term does not affect the appearance of the string ground states. However, it contribuites a torsion density in the string core due to the coupling with the charged particle flux.

The action  of eq.(\ref{acao1}) has a $U(1)' \times U(1)$ symmetry, where the
$U(1)' $ group, associated with the $\phi$-field, is broken by the vacuum and
gives rise to vortices of the Nielsen-Olesen type\cite{Nielsen}

\begin{equation} \begin{array}{ll} \phi = \varphi(r )e^{i\theta},\\
C_{\mu} = \frac{1}{q}[P(r) - 1]\delta^{\theta}_{\mu},
\end{array}\label{vortex1} \end{equation}

\noindent
which are here written in cylindrical coordinates $(t,r,\theta,z)$,
where $r\geq 0$ and $0 \leq \theta < 2 \pi $.  The boundary conditions
for the fields $\varphi(r) $ and $P(r)$ are the same as those of
ordinary cosmic strings\cite{Nielsen}:

\begin{equation}
\begin{array}{ll}
\begin{array}{ll}
\varphi(r) = \eta & r \rightarrow
\infty \\
\varphi(r) =0 & r = 0 \end{array}&
\begin{array}{ll}
P(r) =0 & r \rightarrow \infty \\
P(r) =1 & r= 0.  \end{array}
\end{array} \label{config1}
\end{equation}

The other $U(1)$- symmetry, that we associate to electromagnetism, acts
on the  $\Sigma $-field. This symmetry is not broken by the vacuum;
however, it is broken in the interior of the defect. The $\Sigma
$-field in the string core, where it acquires an expectation value, is
responsible for a bosonic current being carried by the gauge field
$A_{\mu}$. The only non-vanishing components of the gauge fields are
$A_z(r)$ and $A_t(r)$, and the current-carrier phase may be expressed as
$\zeta(z,t) = \omega_1 t - \omega_2z$. Notwithstanding, we focus only
on the magnetic case \cite{Patrick1}. Their configurations are defined
as:

\begin{equation}
\begin{array}{ll}
\Sigma = \sigma(r)e^{i\zeta(z,t)},\\
A_{\mu} = \frac{1}{e}[A(r) - \frac{\partial \zeta(z,t)}{\partial
z}]\delta_{\mu}^{z},
\end{array}
\label{vortex2}
\end{equation}

\noindent
because of the rotational symmetry of the string itself. The fields
responsible for the cosmic string superconductivity have
the following boundary conditions:

\begin{equation}
\begin{array}{ll}
\begin{array}{ll}
\frac{d}{d r}\sigma(r) = 0 &r=0 \\
\sigma(r) = 0 & r \rightarrow \infty \end{array} &
\begin{array}{ll}
A(r) \neq 0 & r \rightarrow
\infty \\
A(r) = 1 & r = 0.
\end{array}
\end{array}
\label{config2}
\end{equation}

Let us consider a SCS in a cylindrical coordinate system $(t,r,\theta,z)$,
so that $r \geq 0$ and $0 \leq \theta < 2 \pi $ with the metric defined in
these coordinates as:

\begin{equation} ds^2 = e^{2(\gamma - \psi)}(-dt^2 + dr^2 ) +
\beta^2e^{-2\psi}d\theta^2 + e^{2\psi}dz^2, \label{metric1}
\end{equation}

\noindent
where $\gamma, \psi$ and $\beta$ depend only on $r$. We can write the
Einstein-Cartan equations in a quasi-Einsteinian form:

\begin{equation} G^{\mu}_{\nu}(\{\}) =8 \pi G (2 \alpha g^{\mu
\alpha}\partial_{\alpha} \Lambda \partial_{\nu} \Lambda  - \alpha
\delta^{\mu}_{\nu} g^{\alpha \beta} \partial_{\alpha } \Lambda \partial_{\beta } \Lambda  + T^{\mu}_{\nu})  = 8\pi G \tilde T^{\mu}_{\nu}, \label{gmg}
\end{equation}

\noindent
where $(\{\})$ stands for Riemannian geometric objects,
$\delta^{\mu}_{\nu} $ and $T^{\mu}_{\nu}$ correspond to the identity
and energy-momentum tensors, respectively. The $\tilde T^{\mu}_{\nu}$
tensor corresponds to an energy-momentum tensor containing the torsion
field.

We have seen that the dependence upon torsion is represented, in the
quasi-Einstenian form, by the $\Lambda$-field that has an equation of
motion given by Eq.(\ref{phi}) below, whose solution shall be presented
subsequently.

The SCS energy-momentum tensor is defined by

\begin{equation}
T^{\mu}_{(scs)\nu} = \frac{2}{\sqrt{g}} \frac{\delta S_m}{\delta g_{\mu \nu}},
\end{equation}

which yields:

\begin{equation}
\begin{array}{ll}
T_{scs \hspace{.1 true cm}t}^t=& -\frac{1}{2} \left\{ e^{2(\psi -
\gamma)}[\varphi'^2 + \sigma'^2] +
\frac{e^{2\psi}}{\beta^2}\varphi^2P^2 + e^{-2\psi} \sigma^2 A^2 +
\right.\\
&\left.+
\frac{e^{2(2\psi-\gamma)}}{\beta^2}(\frac{P'}{q})^2 +
e^{-2\gamma}(\frac{A'}{e})^2 + 2V(\varphi,\sigma, \Lambda)\right\}
\end{array}
\label{tens1}
\end{equation}

\begin{equation}
\begin{array}{ll}
T_{scs \hspace{.1 true cm}r}^r =& \frac{1}{2} \left\{e^{2(\psi -
\gamma)} [\varphi'^2 + \sigma'^2] - \frac{e^{2\psi}}{\beta^2}
\varphi^2P^2 - e^{-2\psi}\sigma^2 A^2 +\right.\\
&\left.+
\frac{e^{2(\psi-\gamma)}}{\beta^2}(\frac{P'}{q})^2 +
e^{-2\gamma}(\frac{A'}{e})^2 - 2V(\varphi,\sigma, \Lambda)\}\right\}
\end{array}
\label{tens2}
\end{equation}

\begin{equation}
\begin{array}{ll}
T_{scs \hspace{.1 true cm}\theta}^{\theta}=& -\frac{1}{2} \left\{
e^{2(\psi - \gamma)} [\varphi'^2 + \sigma'^2] -
\frac{e^{2\psi}}{\beta^2}\varphi^2P^2 + e^{-2\psi }\sigma^2 A^2 +
\right.\\
& -
\left.\frac{e^{2(\psi-\gamma)}}{\beta^2}(\frac{P'}{q})^2 +
e^{-2\gamma}(\frac{A'}{e})^2 + 2V(\varphi,\sigma, \Lambda) \right\}
\end{array}
\label{tens3}
\end{equation}

\begin{equation}
\begin{array}{ll}
T_{scs \hspace{.1 true cm}z}^z =& -\frac{1}{2} \left\{ e^{2(\psi
- \gamma)}[\varphi'^2 + \sigma'^2] +
 \frac{e^{2\psi}}{\beta^2}\varphi^2P^2 - e^{-2\psi }\sigma^2 A^2 +\right.\\
&\left. +
\frac{e^{2(\psi-\gamma)}}{\beta^2}(\frac{P'}{q})^2 -
e^{-2\gamma}(\frac{A'}{e})^2 + 2V(\varphi,\sigma, \Lambda)\right\}.
\end{array}
\label{tens4}
\end{equation}

In the expression of eqs.(\ref{tens1}-\ref{tens4}) only
the usual fields of the string are present.
The Euler-Lagrange equations result from the variation of the
Eq.(\ref{acao1}) together with the conditions for the Nielsen-Olesen
\cite{Nielsen} vortex, Eqs.(\ref{vortex1}-\ref{vortex2}), and yield:

$$
\varphi '' + \frac{1}{r}\varphi ' + \frac{\varphi P^2}{r^2} -
\varphi [\lambda_{\varphi}(\varphi^2 - \eta^2) + 2f_{\varphi \sigma}\sigma^2]=0 $$

$$
\sigma'' + \frac{1}{r}\sigma' + \sigma[A^2 +
(f_{\varphi \sigma}\varphi^2 + \lambda_{\sigma} \sigma^2 -
m_{\sigma}^2 + l^2 \Lambda^2)] =0$$

$$
P'' - \frac{1}{r}P' - q^2 \varphi^2P = 0,
$$

\begin{equation}
A'' + \frac{1}{r}A' - e^2 \sigma^2A =0,
\label{equa1}
\end{equation}

\noindent
while the torsion wave equation  is
given by:

\begin{equation}
\Box_g \Lambda =\frac{l^2}{\alpha} \sigma^2 \Lambda. \label{phi}
\end{equation}

In the equations above, a prime denotes differentiation with respect to the radial
coordinate $r$. The general solution for the SCS will be found in the
weak-field approximation together with junction conditions for the
external metric.

\section{The external solution \zero}

Now, we go on  solving the previous set of equations for an observer
outside the SCS stressed by torsion focusing on the external metric
which satisfies the constraint $r_0\leq r \leq \infty$.  The external
contribution to the energy-momentum of the string reads

\begin{equation}
{\cal T}^{\mu}_{\nu} = \frac{1}{4} g^{\mu \alpha} g^{\beta \rho}
F_{\alpha \beta} F_{\nu \rho } - \delta^{\mu}_{\nu} g^{\sigma \alpha} g^{\beta
\rho} F_{\sigma \beta} F_{\alpha \rho}.\label{stensor1}
\end{equation}

\vspace{.3 true cm}

This tensor is the external energy-momentum tensor of a SCS with no
torsion. If we consider the asymptotic conditions, Eq.(\ref{config1})
 and Eq.(\ref{config2}), we see that the only field that does not vanish
is the $A_\mu$-field. This field is responsible for the conduction of 
the string current. The torsion contribution to the external
energy-momentum tensor is given by

\begin{equation}
{\cal T}_{\nu_{tors} }^{\mu} =  2\alpha g^{\mu \alpha}
\partial_{\alpha}\Lambda \partial_{\nu}\Lambda  - \alpha \delta^{\mu}_{\nu }
g^{\alpha \beta} \partial_{\alpha}\Lambda
\partial_{\beta}\Lambda \label{stensor3} .
\end{equation}

For this configuration, the energy-momentum tensor displays the
following symmetry properties:

\begin{equation} {\cal T}^t_t = -{\cal T}^r_r = {\cal
T}^{\theta}_{\theta}= -{\cal T}^z_z.\label{sim20}  \end{equation}

Then, the only one component of $\Lambda$ in Eq.(\ref{phi}) to be
solved is the $r-$dependent function $\Lambda(r)$. The solution reads:

\begin{equation} \Lambda(r) = \lambda \ln(r/r_0) \label{ssigma}.
\end{equation}

The vacuum solution of Eqs.(\ref{gmg}) are found from the
symmetries Eq.(\ref{sim20}). Hence the solutions of $\beta (r) $ and
$\gamma(r) $ are given by

\begin{equation}
\beta =Br, \hspace{1 true cm}
\gamma = m^2 \ln{r/r_0}.\label{eq2}
\end{equation}

To find the $\psi$-solution, we can use
the condition:

\begin{equation} R = 2 \Lambda'^2 e^{2(\psi - \gamma)}.\label{esca}
\end{equation}

This condition is different from the usual one\cite{Patrick1}
because the scalar curvature $R $ does not vanish, and
opposely it is linked to the torsion-field
 $\Lambda$. Then, this condition has the same form as the one for a
SCS in a scalar-tensor theory \cite{Cris1}. By making use of
solutions (\ref{ssigma}), (\ref{eq2}), we find:

\begin{equation} \psi = n \ln{(r/r_0)} - \ln{\frac{(r/r_0)^{2n} +
k}{(1+k)}}.\label{eq3} \end{equation}

Thus we see from the solutions of the SCCS Eqs.(\ref{eq2},\ref{eq3}),
that there exists a relationship between the parameters $n, \lambda $ and
$m$ given by $n^2 = \lambda^2 + m^2 $.

With the above results, we find that the external metric for the SCS
takes the form:

\begin{equation}
ds^2 =  \left( \frac{r}{r_0} \right)^{-2n} W^2(r) \left[
\left( \frac{r}{r_0}\right)^{2m^2} (-dt^2 +dr^2) + B^2r^2d\theta^2 \right] + \left( \frac{r}{r_0} \right)^{2n}
\frac{1}{W^2(r)} dz^2 \label{m8},
\end{equation}

\noindent
with $W(r) = [(r/r_0)^{2n} + k]/[1+k].$

As it is clear, the external
solution alone does not provide a complete description of the physical
situation. We proceed hereafter to find the junction
conditions to the internal metric,in order to obtain an appropriate
accounting for the nature of the source and its effects on the surrounding space-time.

\section{SCS solution: The weak-field approximation \zero}

Now, let us find the Einstein-Cartan solutions for a SCCS by considering
the weak-field approximation. Thus, the space-time metric may be expanded in terms of a small
parameter $\varepsilon $ about the values $g_{(0)\mu \nu} =
diag(-1,1,1, 1)$, then:

\begin{equation}
\begin{array}{ll}
g_{\mu \nu} = g_{(0) \mu \nu} + \varepsilon h_{\mu
\nu}, \\
\tilde T_{\mu \nu} = \tilde T_{(0) \mu \nu} +
\varepsilon \tilde T_{(1) \mu \nu}.
\end{array}
\end{equation}

The $\tilde T_{_{(0)}\mu \nu}$ tensor corresponds to the
energy-momentum tensor in a space-time with no curvature.
However, torsion is embeeded. $ \tilde T_{_{(1)}\mu \nu}$
represents the part of the energy-momentum tensor containing
curvature and torsion.  Next we proceed to
define some important quantities useful for the analysis to come.

The energy-momentum density  and tension of the thin SCCS are given by:

\begin{equation}
\begin{array}{ll}
U = -2 \pi \int_{0}^{r_0}\tilde T^t_{_{(0)}t} r dr;\\
T = - 2 \pi \int_{0}^{r_0} \tilde T^z_{_{(0)}z} rdr.
\end{array}
\label{U}
\end{equation}

The remaining components follow as

\begin{equation}
\begin{array}{ll}
X = -2 \pi \int_{0}^{r_0} \tilde T^r_{_{(0)}r} r dr;\\
Y = -2 \pi \int_{0}^{r_0} \tilde
T^{\theta}_{_{(0)}\theta} r dr.
\end{array}
\label{X}
\end{equation}

The energy conservation in the weak-field approximation reduces
to

\begin{equation} r\frac{d \tilde T^r_{_{(0)}r}}{dr} = (
\tilde T^{\theta}_{_{(0)}\theta} - \tilde T^r_{_{(0)}r}),\label{bian2}
\end{equation}

\noindent
where $ \tilde T_{_{(0)} \mu \nu}$ represents the trace of the energy-momentum tensor with torsion.

To compute the overall metric, we use the Einstein-Cartan equations in the
quasi-Einsteinian Eq.(\ref{gmg}), where it gets the form $G^{\mu
\nu}(\{\}) = 8 \pi G \tilde T^{\mu \nu}_{_{(0)}}$ in the
weak-field approximation, with the tensor $\tilde T_{_{(0)}\mu \nu}$
(being first order in $G$) containing torsion. After integration, we have:

\begin{equation} \int_0^{r_0} r dr (\tilde T^{\theta}_{_{(0)} \theta}+
\tilde T^r_{_{(0)}r})= r_0^2 \tilde T^r_{_{(0)}r}(r_0) = r_0^2\left[
\frac{A'^2(r_0)}{2 e^2} + \frac{\alpha}{2} \Lambda'^2(r_0)\right] .
\label{ap-1} \end{equation}

To find the internal energy-momentum tensor, it is more convenient to
use Cartesian coordinates\cite{Patrick1}.  For this purpose, we
 calculate the cross-section integrals of $\tilde
T^x_{_{(0)}x}$ and $\tilde T^y_{_{(0)}y}$;  in
Cartesian coordinates, they read as below:

\begin{equation}
\begin{array}{ll}
\tilde T^{x}_{_{(0)}x} = c[\varphi'^2 +
\sigma'^2 + \left(\frac{A'}{e}\right)^2 + \alpha \Lambda'^2] + s\frac{\varphi^2P^2}{r^2} +
\frac{1}{2}\left(\frac{P'}{qr}\right)^2 -\frac{1}{2} \sigma^2 A^2 -
2V \\
\tilde T^{y}_{_{(0)}y} = s [\varphi'^2 +
\sigma'^2 + \left(\frac{A'}{e}\right)^2 + \alpha \Lambda'^2] + c\frac{\varphi^2 P^2}{r^2} +
\frac{1}{2}\left(\frac{P'}{q r}\right)^2 - \frac{1}{2}\sigma^2 A^2
- 2V, \label{tens31}
\end{array}
\end{equation}

\noindent
where $c=\cos^2 \theta -\frac{1}{2}$ and $s=\sin^2 \theta
-\frac{1}{2}$.
We then find:

\begin{equation} \int rdrd\theta \tilde T^x_{_{(0)}x}=\int rdrd\theta
\tilde T^y_{_{(0)}y}= \pi \int r dr[\left(\frac{P'}{q r}\right)^2
- \sigma^2 A^2 - V] = - W.  \end{equation}

Using the fact that
$\tilde T^r_{_{(0)}r} + \tilde
T^{\theta}_{_{(0)}\theta} = \tilde T^x_{_{(0)}x} + \tilde
T^y_{_{(0)}y}$, then we have:

\begin{equation} X + Y = 2 W =-2 \pi r_0^2\left[
\frac{A'^2(r_0)}{e^2} + \alpha \Lambda'^2(r_0)\right] ,
\end{equation}

\noindent
which can be computed by integration of Eq.(\ref{equa1}) so as to give

\begin{equation}
A'(r) = \frac{e J}{\sqrt{2}\pi r}, \hspace{1 true cm}
J= \sqrt{2} \pi e\int_0^{r_0} r dr \sigma^2 A,
\end{equation}

\noindent
where $ J$ is the current density. Thus, the torsion density can be computed by
integration of Eq.(\ref{phi})

\begin{equation}
\Lambda'= \frac{S}{\sqrt{2}\pi \alpha r},\hspace{1 true cm} S =\sqrt{2}\pi  l^2\int^{r_0}_{0} r dr
\sigma^2 \Lambda, \hspace{1 true cm} \label{s2}
\end{equation}

\noindent
where $S$ is the torsion density. With these considerations, we find the
string structure. Then, we obtain

\begin{equation}
W  = -\frac{1}{2\pi }\left(J^2 + \nu S^2\right).
\end{equation}

\noindent
with $\nu = 1/\alpha $.

In addition, we can assume that the string is infinitely thin so that its
stress-energy tensor is given by

\begin{equation} \tilde T^{\mu \nu}_{string} = diag [U , -W, -W,
-T]\delta(x)\delta(y).\label{stensor2} \end{equation}

Is worth  noticing that definitions for both string energy $U$ and
tension $T$, as in equations (\ref{U}), already incorporate
information about the torsion.

By virtue of the presence of the external current, we use the form
Eq.(\ref{stensor2}) for the string energy-momentum tensor as well
as Eq.(\ref{stensor1}) and Eq.(\ref{stensor3}) for the external
energy-momentum tensor in linearized solution to zeroth order in
G. In the sense of distributions we have,

\begin{equation}
\begin{array}{ll}
\nabla^2 ln (r/r_0) = 2 \pi\delta(x) \delta(y)\\
\nabla^2 (\ln (r/r_0))^2 = 2/r^2 \\
\nabla^2(r^2\partial_i \partial_jln(r/r_0))=4\partial_i \partial_j ln(r/r_0).
\end{array}
\end{equation}

The energy-momentum tensor of the string source $\tilde T_{(0)\mu
\nu}$ (Cartesian coordinates) possesses no curvature, which is the
well-known result \cite{Patrick1,Cris1}, but does have torsion which
produces the following energy-momentum tensor

\begin{equation}
\begin{array}{ll}
\tilde T_{(0) tt} = U \delta(x)\delta(y) + \frac{(J^2
+ \nu S^2)}{4\pi}\nabla^2 \left(ln\frac{r}{r_0}\right)^2 ,\\
\tilde T_{(0) zz} = -T \delta(x)\delta(y) + \frac{(J^2
- \nu S^2)}{4\pi}\nabla^2\left(ln\frac{r}{r_0}\right)^2,\\
\tilde T_{(0) ij} = (J^2 +\nu S^2) \delta_{ij}\delta(x)\delta(y)
- \frac{ (J^2 + \nu S^2) }{2 \pi}\partial_i \partial_j ln(r/r_0),
\end{array}
\end{equation}

\noindent
where the trace is given by $\tilde T_{(0)}= - (U + T - J^2 -\nu S^2 )\delta(x)
\delta(y) - \frac{\nu S^2}{2 \pi}
\nabla^2\left(\ln\frac{r}{r_0}\right)^2 $ .

Now, let us find the matching conditions to the external solution. For
this purpose, we shall use the linearized Einstein-Cartan equation in the form

\begin{equation} \nabla^2 h_{\mu\nu} =-16 \pi G (\tilde T_{(0)\mu \nu} -
\frac{1}{2} g_{_{(0)} \mu \nu} \tilde T_{(0)}).\label{ricci1}
\end{equation}

The internal solution to Eq.(\ref{ricci1}) with time-independent source yields:

\begin{equation}
\begin{array}{ll}
h_{tt} =-4 G [J^2(\ln(r/r_0))^2 + (U-T+J^2 + \nu S^2 )\ln(r/r_0)] \\
h_{zz} =-4 G [J^2 \ln(r/r_0))^2 + ( U-T-J^2 - \nu S^2 )\ln(r/r_0)] \\
h_{ij} = -\left[2 G (J^2 + \nu S^2) r^2 \partial_i
\partial_j  - 4 G \delta_{ij} (U+T+J^2 + \nu S^2) \ln(r/r_0)\right] +
S^2\left(\ln \frac{r}{r_0}\right)^2.
\end{array}
\end{equation}

This corresponds to the solution found in Cartesian coordinates. We note
that the torsion appears explicitly in the transverse components of the
metric. To analyze the solution for the with junction condition to the external
 metric, let us transform it back into cylindrical coordinates.

\section{The Matching Conditions \zero}

It is possible to find the junction conditions for
the external solution \cite{Kopczynski}. In the
case of a space-time with torsion, we may find
the matching conditions using the fact
 that $[\{^\alpha_{\mu\nu}\}]_{_{r=r_0}}^{(+)}=
[\{^\alpha_{\mu\nu}\}]_{_{r=r_0}}^{(-)}$, and the metricity constraint
$[\nabla_{\rho}g_{\mu \nu}]_{_{r=r_0}}^{+}= [\nabla_{\rho}g_{\mu
\nu}]_{_{r=r_0}}^{-}= 0$, to get the continuity conditions

\begin{eqnarray}
&[g_{\mu \nu}]_{_{r=r_0}}^{(-)} =
[g_{\mu \nu}]_{_{r=r_0}}^{(+)}, \nonumber \\
&[\frac{\partial g_{\mu
\nu}}{\partial x^{\alpha }}]_{_{r=r_0}}^{(+)} + 2 [g_{\alpha \rho}
K_{(\mu \nu)}^{\hspace{.3 true cm} \rho}]_{_{r=r_0}}^{(+)} =
[\frac{\partial g_{\mu \nu}}{\partial x^{\alpha}}]_{_{r=r_0}}^{(-)} +
2 [g_{\alpha \rho} K_{(\mu \nu)}^{\hspace{.3 true cm}
\rho}]_{_{r=r_0}}^{(-)}, \label{junc1}
\end{eqnarray}

\noindent
where $(-)$ represents the internal region and $(+)$ corresponds the
external region around $r = r_0$. In analyzing the junction conditions
we notice that the contortion contributions do not appear neither in
the internal nor in the external regions\cite{Kopczynski,Volterra}.

To match our solution with the external metric, we used the metric in
cylindrical coordinates,
 which is obtained from the coordinate transformations:

\begin{equation} r^2 \partial_i \partial_j ln(r/r_0)dx^idx^j =
r^2d\theta^2 - dr^2, \end{equation}

\noindent
Unfortunately, for this goal we cannot use the metric
the way it stands. Therefore, we have to change the
radial coordinate to $\rho$, using the constraint (symmetry) $g_{\rho
\rho }= -g_{tt}$, to have, to first order in $G$,

\begin{equation}
\rho=r[1+ a_1-a_2\ln(r/r_0) - a_3(\ln(r/r_0)^2].
\end{equation}

In this case, we have $a_1=G(4U + J^2 + \nu S^2 )$, $a_2= 4GU$ and $a_3=-2G(J^2+ \nu S^2)$,
which corresponds to
the magnetic configuration of the string fields\cite{Patrick1}. The transformed metric yields:

\begin{equation}
\begin{array}{cc}
g_{tt} = -\{1 + 4G[J^2 (\ln(\rho/r_0))^2 + (U - T +
J^2 + \nu S^2)\ln(\rho/r_0)]\} = -g_{\rho \rho}\\
g_{zz} = \{1 - 4G[ (J^2 + \nu S^2)(\ln(\rho/r_0))^2 + (U - T +
J^2- \nu S^2)\ln(\rho/r_0)]\}\\
g_{\theta \theta} = \rho^2\{1 - 8G(U + \frac{(J^2 + \nu S^2)}{2}) +
 4G(U - T - J^2 - \nu S^2)\ln(\rho/r_0)] + 4GJ^2(\ln(\rho/r_0))^2\}\label{m5}.
\end{array}
\end{equation}

Now, we are ready to find the external parameters $B $, $n$ and $m$ as
functions of the source structure. If we consider the junction of
the Eq.(\ref{junc1}),  after the linearization, and using
the limit $|n\ln(\rho/r_0)|<<1$, we have:

\begin{equation}
\begin{array}{ll}
n\left(\frac{1-k}{1+k}\right) = 2G\left( U - T - J^2 - \nu S^2
\right)\\
B^2= 1-8G\left(U + \frac{(J^2 + \nu S^2)}{2}\right)\\
m^2= 4G(J^2 + \nu S^2).
\end{array}
\end{equation}

\noindent
Using the derivative of the expression Eq.(\ref{ssigma}) and the Eq.(\ref{s2}), we arrive at

\begin{equation}
\lambda = \tilde G S.
\end{equation}

\noindent
where $\tilde G = \frac{\nu}{\sqrt{2}\pi }$.
This expression completes the derivation of the full metric
components.  In analyzing the metric of the SCS with torsion, we note
that the contribution of torsion appears in the $\theta \theta$-metric
component, which is important for astrophysical applications such as
gravitational lensing studies, because this component is linked to the
deficit angle.

\section{Screw effect on the propagation  of particles and light \zero}

In this section, we analyze the metric around a screwed
superconducting cosmic string. Let us investigate the deficit angle.
If we consider the metric
Eqs.(\ref{m5}), projected into the space-time perpendicular
to the string, i. e., $dz=0$, then we have:

\begin{equation} ds^2_{\perp}= (1-h_{tt})[-dt^2 + dr^2 + (1- b)
r^2d\theta^2], \end{equation}

with $h_{tt}$ given by

\begin{equation}
h_{tt} = -4G[J^2 (\ln(\rho/r_0))^2 + (U - T +
J^2 + \nu S^2)\ln(\rho/r_0)]\}\label{htt}
\end{equation}

Then, to first order in $G$, the deficit angle gets:

\begin{equation}
\delta \theta = b\pi =8\pi G\{U+\frac{J^2(1+ \ln(\rho/r_0))}{2} +
\frac{1}{2}\nu S^2(1 - 2\ln(\rho/r_0))\}.\label{angle}
\end{equation}

We know that, when the string possesses current, there appear
gravitational forces.  We shall consider the effect that torsion
plays on the gravitational force generated by SCS on a particle
moving around the defect, assumed here the has partide no charge. We consider
the particle speed $|{\bf v}| \leq 1 $, condition under which
the geodesic equation becomes:

\begin{equation}
\frac{d^2x^i}{d\tau^2} + \Gamma_{tt}^{i} =0,
\end{equation}

\noindent
where $i$ is the spatial coordinate, and the connection can be written as
in Eq.(\ref{kont1})). The gravitational acceleration  around the string gets the form

\begin{equation}
{{a}} = \frac{1}{2}( \nabla h_{tt} - \tilde G \frac{S}{\rho}),
\end{equation}

\noindent
with $g_{tt}= -1- h_{tt} $ in Eq.(\ref{m5}).
We also note that the gravitational pull is related to the $h_{tt}$ component
that has  explicit dependence on the torsion, as shown in Eq.(\ref{htt}).
From the last equation, the  force the SCCS exerts on a test particle can
be explicitly written as

\begin{equation}
f = -\frac{m}{\rho} \left[ 2GJ^2\left(1 + \frac{(U-
T + \nu S^2 )}{J^2} + 2\ln(\rho/r_0) \right) + \tilde G S\right]. \label{force}
\end{equation}

\noindent
where $m$ is the mass of the particle.

Let us  consider the deflection of particles moving in the geometry of the
string. For that, we work with the metric in Cartesian coordinates,

\begin{equation}
ds^2 = (1 - h_{tt})[dt^2 -dx^2 -dy^2],\label{lin2}
\end{equation}

\noindent
with $h_{tt}$ given by (\ref{htt}), where for simplicity, we
consider $d_z=0$. The linearized geodesic equations in this metric
takes the form

\begin{equation}
2\ddot{x} = -(1-\dot{x}^2-\dot{y}^2)\partial_xh_{tt} +(1-\dot{y}^2)\partial_x\phi
\end{equation}

\begin{equation}
2\ddot{y} = -(1-\dot{x}^2-\dot{y}^2)\partial_yh_{tt} +(1-\dot{x}^2)\partial_y\phi, \label{y}
\end{equation}

\noindent
with dots referring to derivatives with respect to t. To calculate
the transverse velocity  the particles acquire after passing by the
string,  we consider that the particles are flowing with initial
velocity, $v$, and we can  integrate over the unperturbed
trajectory, $x=vt$ and $y=y_0$

The result is the velocity
impulse in the $y$ direction after the string
has passed. Then, particles enter the wake with
a transverse velocity:

\begin{equation}
v_t = 4\pi  G(U + T + I^2 + \nu S^2)v \gamma + \frac{4\pi \tilde G S}{  v
\gamma}.\label{vy}
\end{equation}

The first term is the usual velocity impulse of the particles due to
the deficit angle; the additional term is nothing but the torsion contribution.

A quick glance at the last equation may illustrate  the
essential role torsion eventually plays in the context of cosmic string.
If torsion is present a new attractive  gravitational
force acts on test particles. 
According to Pogosian and Vachaspati\cite{Pogosian},
it is possible that wiggles may impart
significant longitudinal velocities to string segments; a
characteristic torsion also imprints the same effect, as we can verify
with the help of (\ref{vy}). In this statement, we consider $\nu$ arbitrary,
but, in the case  $\nu$ is of the order of $G$,  the
contribution of the angular deficit is of second order in $G$, hence negligeable;
the second contribution is of order of  $G$,
and may therefore have non-negligeable effects.

\section{Conclusions and Remarks}

It is possible that torsion may have had a physically relevant role
during the early stages of the Universe's evolution. Along these
 lines, torsion fields may be regarded as  potential sources for 
dynamical stresses which, when coupled to other fundamental fields 
 (the gravitational field, for instance ), might have had a significant
 contribution during the phase transitions leading to formation of
 topological defects, such as the SCSs  we focused on here.
It therefore seems a  crucial issue to investigate basic models and
scenarios involving  cosmic defects within a context where torsion
degrees of freedom are suitably accounted for.
 We  showed that torsion  has a small, but non-negligeable, contribution
 to the geodesic equation obtained from the contortion term.

From a physical point
of view, this contribution is responsible for the appearance of an
additional  attractive force acting on a massive test-particle, as
explained in Section 6; this attractive force is  important to the
accretion of matter by the wake; in such a situation, the screwed cosmic
string behaves very much like wiggly cosmic strings,  assuming, of course, the
validity of the Pogosian and Vachaspati conclusion \cite{Pogosian}.
 
To incorporate the Pogosian and Vachaspati
statement\cite{Pogosian}, we propose that their straight strings
with small-scale structures ($wiggles$) may  resemble the strings
endowed with torsion  in our picture (screwed strings). In so
doing, we postulate that the small-scale structures existing in
wiggly strings can be approximately scaled to the geometrical deformation
torsion produces on  ordinary strings. Admitting this
premise leads us to the idea that the primordial spectrum of
perturbations in the CMBR, as observed by COBE, may reasonably
be reproduced if one uses the freedom in the parameter-space of
numerical models of structure formation based on wiggly strings.

The shape of the matter (radiation) power spectrum can be obtained
by following the evolution of a network of long ordinary straight
strings interacting with the universe matter (radiation) content.
A string evolves in such a way that its characteristic curvature
radius at time $t$ is $\sim t$ (the reader is addressed to
Ref.\cite{SV84}, and references therein, for an exhaustive
discussion on this  subject). Each string moves with typical speed
$v $ through matter. The translational motion of a string creates
a wedge-shaped wake. Once the wake  forms, particles fall into it
with transversal velocity, $v_t \sim 8 \pi G \mu v \gamma $ (where
$\mu$ is the linear mass density of the string, $\gamma = (1-
 v^2)^{-1/2}$), towards the plane behind the string  \cite{Pogosian}.
Test particles that do not collide travel a distance shorter than the wake
 width: $v_t^2/2a_w \sim  4 \pi G \mu t$.
 
Now that we have already studied the wake formation, we are in a
position to analyze the wake generating by screwed superconducting
cosmic string.
 The study of the effects of a cosmic string passing through matter
is of great importance to understand the current organization of
matter in the Universe and in this context many authors have
 already considered this problem in General Realativity.\cite{Vachaspati}
 
Massless particles (such as photons) will be deflected by an angle
$\delta \theta = 4 \pi G(U+T+I^2)$ in the case where $\tilde G
\sim G$. From the observational point of view, it would be
impossible to distinguish a screwed string from its General
Relativity partner, just by considering effects based on deflection
of light (i.e., double image effect, for instance). On the other
hand, trajectories of massive particles will be affected by the
torsion coupling (which generates by torsioned space-time).
These results are  compatible with the Kleinart conclusions  already
shown in the Introduction.
 
 If the string is moving with normal velocity, $v$, through matter,
a transversal velocity appears that is given by Eq.(\ref{vy}). We
can see that there exists a new contribution to the transversal
velocity $v_t = \frac{4 \pi \tilde G S}{v \gamma}$ given
by torsion.
 We saw  that the propagation
of photons is unaffected by a screwed superconducting cosmic string,
and is only affected by the angular deficit. This result shows us that the
effect of torsion on massive particles is qualitatively
different from its effect on radiation; this aspect becomes  especially
relevant when calculating CMBR-anisotropy and the power spectrum
as wiggly cosmic strings. One expects  that this feature could
help to partially by-pass the current difficulties in reconciling the COBE
normalized matter power spectrum with the observational data in
the cosmic string model.

{\bf Acknowledgments:} The author is indebted to J. H. Mosquera-Cuesta and
L.C.Garcia-de-Andrade for participation at an early stage of this work.
Thanks are also due to J. A. Helayel-Neto and V. B. Bezerra for long
discussions, suggestions and careful criticisms on the original manuscript  of this paper. C.N.Pq.-
Brazil is acknowledge for the Graduate fellowship.


\begin{thebibliography}{99}

\bibitem{Vilenkin} A.Vilenkin, Phys. Rev. {\bf D 23},852 , (1981);
W.A.Hiscock, Phys. Rev. {\bf D 31}, 3288, (1985); J.R.GottIII,
Astrophys. J. {\bf 288}, 422, (1985); D. Garfinkel, Phys. Rev.
{\bf D 32} 1323, (1985).

\bibitem{Kibble} T.W. Kibble, J. of Phys. {\bf A9}, 1387 (1976).



\bibitem{avelino98} P. P. Avelino et al., Phys. Rev. Letts. 81, 2008 (1998).




\bibitem{smoot99}G. F. Smoot, Lectures at D.Chalonge School on
Large-Scale Structure of the Universe, Erice, Dic 12-17 (1999).



\bibitem{Pogosian}L. Pogosian and T. Vachaspati, Phys. Rev. D 60, 083504 (1999).




\bibitem{Brandenberger} J. Magueijo and R.H. Brandenberger,
astro-ph/0002030, (2000); T.W.B. Kibble, Phys. Rep {\bf 67}, 183,
(1980).

\bibitem{brandeb93} R. H. Brandenberger, A. T. Sornborger and M.
 Trodden, Phys. Rev. D. 48, 940 (1993); V. Berezinshy, B. Hnatyk and A. Vilenkin,
astro-ph/0001213 (2000).

\bibitem{HD00} H. J. Mosquera Cuesta and D. Mor\'ejon Gonz\'alez, to appear in Phys. Letts. B. (2000).


\bibitem{Allen94} B.Allen and P.Casper, Phys. Rev. {\bf D 51}, 1546
(1995), gr-qc/9407023; Phys. Rev. {\bf D 50}, 2496 (1994),
gr-qc/9405005.

\bibitem{Hawking} S.W.Hawking and S.F.Ross, Phys.Rev.Lett. 75, 3382 (1995), gr-qc/9506020 (1995);
Audretsch and A.Economou, Phys. Rev. {\bf D 44},3774 (1991).

\bibitem{Witten1} E. Witten, Nucl.Phys. {\bf B249}, 557 (1985).


\bibitem{Patrick1} P. Peter and D. Puy, Phys. Rev. D 48, 5546 (1993).

\bibitem{Jackiw} R. Jackiw and P. Rossi, Nucl. Phys {\bf B 190}, 681, (1981).

\bibitem{Davis99} S.C.Davis, Int. J. Theor. Phys. 38, 2889
(1999), hep-ph/9901417 (1999);  S.C.Davis {\it et al.}, Phys. Rev.
{\bf D 62}, 043503, (2000), hep-ph/9912356 (1999).

\bibitem{Sabbata1} V. De Sabbata, IL Nuovo Cimento, {\bf 107 A}, 363, (1994).


\bibitem{Yishi} Y.Duan, G. Yang and Y. Jiang, Helv. Phys. Acta {\bf 70}, 565, (1997)



\bibitem{Deser1} S. Deser and A.N. Redlich, Phys. Lett. 176B (1986) 350;

\bibitem{Boulware} D.G. Boulware and S. Dese, Phys. Rev. Lett. 55 (1986) 2656
and Phys. Lett. 175B (1986) 409;

\bibitem{Deser2} S. Deser, "Gravity from Strings", in Unification of Fundamental
Interactions Proc. of Nobel Symposium 67,
ed. by L. Brink, R. Marnelius, J.S. Nilsson, P. Salomonson and
B.-S. Skagerstam, Marstrand - Sweden, June 1986.


\bibitem{Shapiro2} I. L. Shapiro, hep-th/0103093 to appear in Phys. Rept.





\bibitem{Ogino} S. Ogino, Prog. Theor. Phys. {\bf 73}, 84, (1985).

\bibitem{Ross} D. K. Ross, Int. J. Theor. Phys. {\bf 28}, 1333, (1989)

\bibitem{Sabbata3} V. De Sabbata, IL Nuovo Cimento, {\bf A107} 363, (1994)  


\bibitem{Soleng92} H.H. Soleng, Gen. Rel. Grav. {\bf 24}, 111, (1992).




\bibitem{Letelier1} P.S.Letelier, Class. Quant. Grav.{\bf 12}, 471
(1995); P.S.Letelier, Class. Quant. Grav. {\bf12}, 2224, (1995).








\bibitem{Kleinert} H.Kleinert, Phys.Lett.{ \bf B440} 283 (1998).


\bibitem{Kleinert1} H.Kleinert, Gen. Rel. Grav. { \bf 32} 769 (2000).

\bibitem{Kleinert2} H.Kleinert, Gen. Rel. Grav. { \bf 32} 1271 (2000).












\bibitem{Gaspperini} V. de Sabbata and M.Gasperini, ``Introduction to Gravitation ", World Scientific Publishing (1985)

\bibitem{Hehl} F.W.Hehl et al., Rev. Mod. Phys. {\bf 48}, 393 (1976).




\bibitem{Nielsen} H.B Nielsen and P. Olesen, Nucl.Phys. {\bf 61}, 45,
(1973).

\bibitem{Cris1} C.N.Ferreira, M.E.X. Guimar\~aes and J.A.Helayel-Neto, Nucl.Phys. {\bf B 581}, 165, (2000).


\bibitem{Kopczynski} W.Arkuszewski {\it et al.},
Commun.Math.Phys. {\bf 45}, 183 (1975).



\bibitem{Volterra} R.A.Puntigam and H.H.Soleng, Class. Quantum Grav.
{\bf 14}, 1129 (1997).




\bibitem{SV84}J. Silk and A. Vilenkin, Phys. Rev. Letts. 53, 1700 (1984).





\bibitem{Vachaspati} T. Vachaspati, Phys. Rev. Lett. {\bf 57}, 1655, (1986);
Phys. Rev. {\bf D 45}, 3487, (1992); A. Stebbins , S. Veraraghavan,
R. H. Brandenberger, J. Silk and N. Turok, Astrophys. J. {\bf 1}, 322 (1987).


\end{thebibliography}
\end{document}